\documentclass[prd,nofootinbib,showkeys]{revtex4}
\usepackage{amsmath,graphicx}

\begin{document}

\title{Information on the structure of the rho meson from the pion form-factor}
\author{S. Leupold}
\affiliation{Institut f\"ur Theoretische Physik, Universit\"at
Giessen, Germany}

\begin{abstract}
The electromagnetic pion form-factor is calculated in a Bethe-Salpeter approach 
which accounts for 
pion rescattering. In the scattering kernel the pion-pion contact interaction
from lowest-order chiral perturbation theory is considered together with an optional
vector meson in the s-channel. Correspondingly the virtual photon couples to a two-pion
state and optionally to the vector meson. It is shown that for reasonable ranges of input 
parameters the experimentally observed pion form-factor cannot be described by the 
iterated pion-pion contact interaction alone, i.e.\ without an elementary vector meson. 
The inclusion of an elementary vector meson allows for an excellent description.
This completes a recent study 
(``Information on the structure of the $a_1$ from tau decay'')
where it has been shown that the $a_1$ meson can be 
well understood as a rescattering process of $\rho$ meson and pion. Here it is demonstrated
that within the same formalism the $\rho$ meson cannot be understood as a pion-pion
rescattering process. This suggests that the chiral partners $a_1$ and $\rho$ are
not only different in mass, but also different in nature.
\end{abstract}

\maketitle

\section{Introduction and summary}
\label{sec:intro}

The $\rho$ meson and the $a_1$ meson are chiral partners in the following sense: As deduced, 
e.g., from tau decays \cite{Schael:2005am} the $\rho$/$a_1$ meson
couples strongly to the isovector current of vector/axial-vector type. On the other hand,
vector and axial-vector currents are related by chiral transformations. The fact that the
masses of the $\rho$ and $a_1$ meson are very different is explained by the spontaneous breaking 
of chiral symmetry. 
According to common lore the $\rho$ and the $a_1$ meson are both quark-antiquark states. In
this quark-model picture the two mesons have the same nature, but differ in mass due to
chiral symmetry breaking. 

Recently, this picture has been cast into doubt. As has been
shown in \cite{Lutz:2003fm} the spectrum of lowest-lying axial-vector mesons --- including
the $a_1$ meson --- can be understood as being dynamically generated from coupled-channel
scattering processes of Goldstone bosons on vector mesons. Concerning the $a_1$ meson
this has been reconfirmed in a technically somewhat different 
approach in \cite{Roca:2005nm}.
In these works \cite{Lutz:2003fm,Roca:2005nm} axial-vector meson masses have been 
deduced from
the scattering amplitudes. The latter, however, are not observable quantities since
the vector mesons are not asymptotic states.\footnote{Strictly speaking also the Goldstone
bosons are not asymptotic states, but, e.g., the pions are stable with respect to the 
strong interaction. Therefore the pions in practice live long enough to allow for pion 
beams and for pions reaching the detectors. In contrast, vector mesons only show up
in scattering phase shifts and not as scattering partners.} 

An analysis of a real observable, namely the semi-hadronic tau decay into three
pions, has been studied in \cite{Wagner:2008gz,Wagner:2007wy}. In this decay process the 
$a_1$ meson appears as a prominent bump in the spectrum. As demonstrated in 
\cite{Wagner:2008gz,Wagner:2007wy} this bump can be well explained by a $\pi$-$\rho$ 
rescattering
process without the need for an elementary (quark-antiquark) axial-vector meson. 
This immediately
raises the question whether also the $\rho$ meson, as the chiral partner of the $a_1$ meson,
can be seen as a meson-meson rescattering process 
without the need for an elementary (quark-antiquark) vector meson. 
In principle, such a question has
already been addressed in \cite{Oller:1998zr}. 
Based on the $N/D$ method, it has been shown 
that pion-pion rescattering 
without an elementary vector meson is not
sufficient to describe the elastic pion-scattering phase shifts. 

In the present paper
we will add three aspects to the analysis of the $\rho$ meson. 
First, to obtain a consistent picture with
our $a_1$ analysis \cite{Wagner:2008gz,Wagner:2007wy}, we will use a 
Bethe-Salpeter approach, 
i.e.\ the very same method which explained the $a_1$ bump in the tau decay data by 
meson-meson rescattering without an elementary 
axial-vector meson \cite{Wagner:2008gz,Wagner:2007wy}. 
Second, we do not concentrate on the pion-scattering phase shift but rather on the 
electromagnetic pion form-factor as this probes the 
isovector--vector channel
just like the tau decay into three pions probes the isovector--axial-vector channel. 
As an intermediate step we will also obtain
the elastic pion-scattering phase shift for the p-wave isovector channel. 

We will study here two scenarios to describe the pion form-factor: one with and one
without an elementary vector meson. In both cases we take into account the 
pion-pion and pion-photon interactions from lowest-order chiral perturbation theory. 
It is an important aspect that these interactions are fixed model-independently
by chiral \cite{Lutz:2003fm,Roca:2005nm,Oller:1998zr} or charge symmetry, respectively.
In full qualitative agreement with \cite{Oller:1998zr} we will find that an 
elementary vector meson is
needed to describe the pion form-factor. 
Thus, together with our previous analysis of the $a_1$ 
meson \cite{Wagner:2008gz,Wagner:2007wy}
it is suggestive that $\rho$ and $a_1$  are not
only different in mass, but also in nature 
($a_1$ = dominantly $\pi$-$\rho$ state, $\rho$ = dominantly quark-antiquark state). 
Such a finding is in line with the hadrogenesis
conjecture \cite{Lutz:2003fm,Lutz:2005ip,Lutz:2008km} which implies 
that the lowest-lying pseudoscalar and
vector mesons (and in the baryon sector the nucleon octet and the Delta decuplet)
are dominantly quark-antiquark (three-quark) states while other low-lying hadrons are 
dynamically generated.

There is a third, technical, aspect why we present this analysis in spite of the fact that
the qualitative outcome is fully in line with the findings of \cite{Oller:1998zr}: 
Typically the 
spectral information of a resonance is determined in a Dyson-Schwinger approach, i.e.\ by
determining a one-body quantity, the self energy. In contrast, a Bethe-Salpeter approach
starts from the objects which form the resonance, i.e.\ from a two-body quantity, the
scattering amplitude. It is interesting to work out the connections between the two
approaches, in particular for the non-trivial case where there is not only the resonance
in the scattering kernel, but also an additional non-resonant, here point-like, 
interaction.

For the pion rescattering we will restrict ourselves in the following 
to the elastic channel, as will be 
discussed in more detail below. In principle, it is straightforward to include the
$K \bar K$ channel as intermediate states. However, there is a second channel
with about the same threshold, namely the $\pi\omega$ channel which then should also be 
included. It can also be seen as one, presumably 
important, representative of the four-pion channel. The inclusion of the $\pi\omega$ 
channel is more demanding since the scattering kernel of $\pi \pi \to \omega \pi$ is not
in the realm of chiral perturbation theory. The 
power counting for pseudoscalar and vector mesons as developed in 
\cite{Lutz:2008km,Leupold:2008bp} can be used for such an extension. However,
the interactions which play a role there are not restricted to point interactions and
s-channel resonance formation while they are in our present approach as we will
discuss in more detail below. Rather t- and u-channel exchange diagrams come into
play there which calls for a proper treatment of the appearing left-hand 
cuts within a resummation scheme like the Bethe-Salpeter approach 
(cf.\ also the discussion in \cite{Oller:1998zr,pelaez02}). Without a proper handling
of the left-hand cuts analyticity is spoiled and statements concerning the possibility
to dynamically generate a state become questionable.
Such an extension is beyond the scope of
the present work. We therefore restrict ourselves to elastic pion rescattering and leave
the inclusion of the channels $K \bar K$ and $\pi\omega$ for future work.

The paper is structured in the following way: In the next section we present the formalism
to calculate the pion form-factor. After some general considerations we will formulate
the Bethe-Salpeter equation for pion rescattering in subsection \ref{sec:scatt}.
The pion form-factor is addressed in subsection \ref{sec:BS}. Subsection \ref{sec:spec}
is devoted to a comparison of the Dyson-Schwinger and the Bethe-Salpeter equation.
In section \ref{sec:results} the results are presented and compared to the experimental
data for the pion form-factor and the pion-scattering phase shift. The two subsections 
\ref{sec:wo} and \ref{sec:with} concern the two scenarios
without and with an elementary vector meson. Finally an appendix is added for some
technical details.

\section{Formalism}
\label{sec:formalism}

\subsection{General considerations}
\label{sec:gen}

The electromagnetic pion form-factor is defined via
\begin{equation}
\langle \pi^+(p') \vert j^\mu_{\rm em} \vert \pi^+(p) \rangle = (p+p')^\mu F_\pi(q^2) 
\label{eq:defpff}
\end{equation}
with $q=p-p'$ and the electromagnetic current $j^\mu_{\rm em}$. The normalization is 
chosen such that the form factor is unity for the pure QED process where only the charge 
of the pion is probed.
The time-like region of the pion form-factor is accessible by the reaction 
$e^+ e^- \to \pi^+ \pi^-$ \cite{Barkov:1985ac,Akhmetshin:2001ig,Akhmetshin:2003zn}. 
The pion form-factor has been studied by many groups; 
see, e.g., \cite{Guerrero:1997ku,Nieves:1999bx,Pich:2001pj,Oller:2000ug} 
and references therein. 
However, a Bethe-Salpeter approach
has rarely been used and the question whether one can understand the pion form-factor
by pure rescattering of pions without an elementary vector meson has not been addressed.
For a Bethe-Salpeter approach, i.e.\
including rescattering one needs at least the scattering kernels for the following
reactions: $e^+ e^- \to \pi^+ \pi^-$ and $\pi^+ \pi^- \to \pi^+ \pi^-$. In principle,
one also needs the elastic channel $e^+ e^- \to e^+ e^-$, but as a non-hadronic channel
its contribution is negligibly small (cf.\ also the appendix). 
One can also imagine to consider other
intermediate states like $K\bar K$ or $4\pi$. In the following we restrict ourselves to 
center-of-mass energies below 1 GeV. The threshold for $K\bar K$ production
lies at about 1 GeV and it opens up as a p-wave. Thus it should be reasonable to neglect
this channel. If one included $K\bar K$, one should presumably also consider $\pi\omega$
which is a representative of the four-pion channel. The nominal threshold for $\pi\omega$
is at about 900 MeV. For uncorrelated pions the $4\pi$ threshold is even below 600 MeV.
Experimentally, however, it turns out that the four-pion channel is not very important
below 1 GeV \cite{Dolinsky:1991vq}. 
Therefore, we keep things as simple as possible and consider only two-pion
intermediate states for the reaction of interest, $e^+ e^- \to \pi^+ \pi^-$.

Finally, a comment is in order concerning the three-pion intermediate state. This state
is forbidden by G-parity, but since isospin is not an exact symmetry, such an intermediate
state is possible. In particular the three pions might be correlated to an $\omega$ meson.
Thus the small probability to violate isospin can be overcompensated by the large 
probability to create a sharp resonance. Indeed, one observes the $\omega$ meson in the
reaction $e^+ e^- \to \pi^+ \pi^-$ (``$\rho$-$\omega$ mixing''). However, if one stays outside
of the isolated $\omega$ peak there is no further visible trace of the three-pion intermediate 
state. The present work is not concerned with the isospin violating mixing to the $\omega$
meson. It is well known how to include this mixing on a 
phenomenological level \cite{klingl1}.
For the present work we ignore this aspect and demand exact isospin symmetry.

For the construction of the scattering kernels we follow the logic 
of \cite{Wagner:2008gz,Wagner:2007wy}: 
If there is an elementary resonance in the kinematical
region of interest we include the corresponding s-channel diagram in the kernel. All other 
interactions are smooth functions of the center-of-mass energy $\sqrt{s}$. We approximate
these non-resonant interactions by the respective lowest-order term of chiral perturbation 
theory \cite{Gasser:1983yg,Gasser:1984gg,Scherer:2002tk}. 
Of course, one has to make sure that there is
no double counting in this procedure. We will come back to this point below.

For our case of interest we have the following contributions to the kernels:
\begin{eqnarray}
  k_{e^+ e^- \to \pi^+ \pi^-} & = & k^{\rm QED}_{e^+ e^- \to \gamma^* \to \pi^+ \pi^-} 
  + k^{\rm res}_{e^+ e^- \to \gamma^* \to \rho^0 \to \pi^+ \pi^-}  \,,
  \label{eq:KFF}
  \\ 
  k_{\pi^+ \pi^- \to \pi^+ \pi^-} & = & k^{\rm point}_{\pi^+ \pi^- \to \pi^+ \pi^-} 
  + k^{\rm res}_{\pi^+ \pi^- \to \rho^0 \to \pi^+ \pi^-} \,.
  \label{eq:Kpipi}
\end{eqnarray}
The non-resonant contributions $k^{\rm QED}$ and $k^{\rm point}$ emerge from chiral 
perturbation theory in lowest, i.e.\ second, 
order \cite{Gasser:1983yg,Gasser:1984gg,Scherer:2002tk}. 
$k^{\rm QED}$ is just the QED-type
process where the virtual photon couples to the charge of the pion. $k^{\rm point}$ is
the pion four-point interaction of the non-linear sigma model. 

As already announced we study two scenarios: One with and one without an elementary
vector meson. In the latter case we set $k^{\rm res}_{\ldots} =0$ in 
(\ref{eq:KFF}) and (\ref{eq:Kpipi}). For the former case we follow 
\cite{eckgas,Lutz:2008km} and use the tensor realization of the vector mesons.
If the vector mesons were integrated out for low energies, they would only contribute 
to the fourth-order Lagrangian of chiral perturbation theory, not to the second-order 
one \cite{eckgas,Ecker:1989yg}. Thus there is no double counting in 
(\ref{eq:KFF}) and (\ref{eq:Kpipi}) on the level of the kernels.

The Bethe-Salpeter equation iterates the kernels to infinite order. In that way loops
emerge which require renormalization. In general, the renormalization of resummed series
is not as clear-cut as the renormalization of a perturbation theory. Ambiguities arise
which can also influence the issue of double counting \cite{Hyodo:2008xr}. 
We will come back
to that point below when discussing the Bethe-Salpeter equation in more detail.

\subsection{Pion-scattering amplitude}
\label{sec:scatt}

We discuss in the following the more general scenario which includes an elementary
vector meson. The other scenario can easily be obtained by putting the appropriate
coupling constants ($e_V$ and $h_P$, see below) to zero.
We specify first the scattering amplitude for elastic pion-pion scattering:
In the center-of-mass system the Feynman scattering amplitude ${\cal M}$ is decomposed 
into amplitudes $t_l$ with fixed orbital angular momentum $l$:
\begin{equation}
{\cal M}(s,\cos\theta) = \sum\limits_l (2l+1) \, t_l(s) \, {\rm P}_l(\cos\theta)
\label{eq:l-decomp}
\end{equation}
with the Legendre polynomials ${\rm P}_l$. The phase shift is given by
\begin{equation}
\cot \delta_l = \frac{{\rm Re} \, t_l}{{\rm Im} \, t_l}
\label{eq:l-el-ph-sh}
\end{equation}
and the optical theorem reads
\begin{equation}
{\rm Im} \left(t_l^{-1} \right) = -\frac{p_{\rm cm}}{8\pi\sqrt{s}}
\label{eq:opt-theo}
\end{equation}
with the center-of-mass momentum $p_{\rm cm}$. 

The final process of interest, $e^+ e^- \to \pi^+ \pi^-$, proceeds via a virtual 
photon in the s-channel. Consequently
the orbital angular momentum of the two-pion system is fixed to $l=1$. The isospin
is then fixed to $I=1$ to allow for an overall symmetric two-pion state. 

The scattering amplitude can be decomposed into two-particle reducible and
two-particle irreducible parts. The latter constitute the kernel of the 
Bethe-Salpeter equation while the former are automatically generated by this
equation. As a first step we need an approximation for the kernel. As already
spelled out we use lowest-order chiral perturbation theory plus a (bare)
vector-meson s-channel diagram. Note that the width of the vector meson is
generated by the Bethe-Salpeter equation. Hence both parts are tree-level
contributions. 

With the conventions of \cite{Lutz:2008km} we find
\begin{eqnarray}
k_{\pi^+ \pi^- \to \pi^+ \pi^-} = k_{l=1,I=1}(s) = 
	\frac{2}{3\,f^2} \, p_{\rm cm}^2 \, 
	\left(1-\frac{m_V^2 \, h_P^2}{8\, f^2} 
		\frac{s}{s-m_{\rho,{\rm bare}}^2} \right) \,.
\label{eq:k-scatt-tree}
\end{eqnarray}
Here $f=90\,$MeV denotes the pion-decay constant in the chiral limit and
$m_{\rho,{\rm bare}}$ the mass of the elementary vector meson. The
combination $m_V h_P$ parametrizes the coupling of the vector meson to
the pions. The dimensionful quantity $m_V$ is conveniently chosen as
$m_V = 776 \,$MeV. 
In the following we use the dimensionless quantity $h_P$ as
a free input parameter which will be adjusted to the data for the scenario
including an elementary vector meson. For the alternative scenario, i.e.\ 
where there is no elementary vector meson, we simply set $h_P$ to zero.
Finally we note for completeness: $p_{\rm cm} = \frac12 \sqrt{s-4 m_\pi^2}$
with the pion mass $m_\pi$.

With the formalism of \cite{Lutz:2003fm} the Bethe-Salpeter equation reads
\begin{equation}
t^{-1}_{l=1,I=1}(s) = k^{-1}_{l=1,I=1}(s) - I(s) + {\rm Re} \, I(\mu^2)  \,.
\label{eq:BS}
\end{equation}
Here $I(s)$ is the loop function
\begin{equation}
I(s=p^2) = -i \int \frac{{\rm d}^dq}{(2\pi)^d} \, 
	\frac{1}{[q^2-m_\pi^2 +i\eta]\, [(q-p)^2-m_\pi^2 + i\eta]}  \,,
\label{eq:def-loop}
\end{equation}
regularized by dimensional regularization with $d=4+2\epsilon$.
Several remarks are in order here:
\begin{enumerate}
\item Unitarity (\ref{eq:opt-theo}) is exactly fulfilled by 
  (\ref{eq:BS}) since
  ${\rm Im} \, k = 0$ (two-particle irreducibility) and
  ${\rm Im} \, I(s) = p_{\rm cm}/(8\pi\sqrt{s})$. This is in contrast
  to any perturbation theory which satisfies unitarity only 
  perturbatively, but not exactly. On the other hand, exact unitarity
  is an important requirement in the resonance region \cite{pelaez02}, i.e.\ for
  energies larger than the region of applicability of strict chiral
  perturbation theory. 
\item	Analyticity is also fulfilled by (\ref{eq:BS}) since the kernel
  $k$ as a rational function --- given in (\ref{eq:k-scatt-tree}) ---
  does not have any cuts. This property is actually a necessary requirement
  to write down the Bethe-Salpeter equation in an analytic way as given by
  (\ref{eq:BS}). 
\item	Crossing symmetry is not fulfilled by any Bethe-Salpeter equation
  since processes are iterated in the s-channel, but not in the t- and
  u-channel. As pointed out in \cite{Lutz:2003fm} {\em approximate} crossing
  symmetry can be ensured by a proper choice of the renormalization point
  $\mu$ (see next item).
\item The loop function is renormalized by the replacement
  \begin{equation}
    I(s) \to I(s) - {\rm Re} \, I(\mu^2) = J(s) - {\rm Re} \, J(\mu^2) 
    \label{eq:renorm}
  \end{equation}
  where we have introduced the finite function
  \begin{equation}
    J(s) = \frac{1}{16\pi^2} 
    \left(2 + \sigma(s) \, \log\frac{\sigma(s)-1}{\sigma(s)+1}\right)
    \label{eq:defJ}
  \end{equation}
  and the phase space $\sigma(s) = 2p_{\rm cm}/\sqrt{s}$.
  By the replacement (\ref{eq:renorm}) a new parameter, the
  renormalization point $\mu$, is introduced. Approximate crossing
  symmetry is ensured if the full scattering amplitude $t$ reduces
  to the perturbative amplitude $k$ below and close to 
  threshold \cite{Lutz:2003fm}. For the scenario with an elementary
  vector meson we will follow this requirement and choose 
  \begin{equation}
    \mu \approx m_\pi \,.
    \label{eq:choicemu}
  \end{equation}
  We will study the impact of moderate changes.
  For the scenario without an elementary vector
  meson we will allow for arbitrary changes of the renormalization point. But
  we will keep in mind that a drastic deviation from (\ref{eq:choicemu})
  is physically questionable. As pointed out in \cite{Hyodo:2008xr} an
  improper choice of $\mu$ is even related to the double-counting
  problem raised above. We will come back to that point below when
  we dicuss the results for our two scenarios.
\end{enumerate}

\subsection{Pion form-factor}
\label{sec:BS}

The process $e^+ e^- \to \pi^+ \pi^-$ can be treated within the Bethe-Salpeter
approach as a two-channel problem. A simplification arises, however, if one
treats the electromagnetic processes on a perturbative level. This issue is
worked out in the appendix. The result is
\begin{equation}
	t_{e^+ e^- \to \pi^+ \pi^-}(s) \approx k_{e^+ e^- \to \pi^+ \pi^-}(s) \,
		\left[ 1 + \left(I(s) - {\rm Re} \, I(\tilde \mu^2) \right) \, t_{l=1,I=1}(s) 
		\right]
\label{eq:offdiag}
\end{equation}
with the elastic pion scattering amplitude $t_{l=1,I=1}$ introduced already in
(\ref{eq:BS}). 

It is worth to discuss the renormalization issue,
i.e.\ the replacement $I(s) \to I(s) - {\rm Re} \, I(\tilde \mu^2)$ which led
from (\ref{eq:t12approx}) to (\ref{eq:offdiag}). Note in particular that we have 
introduced a new renormalization point $\tilde \mu$ here which we kept distinct
from the renormalization point $\mu$ of pion rescattering appearing 
in (\ref{eq:BS}).
In principle, the coupled-channel Bethe-Salpeter approach \cite{Lutz:2003fm} 
demands that the renormalization points $\mu$ and $\tilde \mu$ should be
the same:
\begin{equation}
\tilde \mu = \mu \approx m_\pi \,.
\label{eq:same-renorm}
\end{equation}
For the scenario with an elementary vector meson in the kernels we will follow
this demand (\ref{eq:same-renorm}). For the case without an elementary vector
meson we will explore the consequences of a free choice for $\mu$ and 
$\tilde \mu$ independent of each other.

It is interesting to figure out which loop is actually renormalized in 
(\ref{eq:offdiag}): The first contribution to $t_{e^+ e^- \to \pi^+ \pi^-}$
(coming from the $1$) is just the emission of two pions without rescattering.
The second contribution is the process where the pions rescatter. Thus the
loop function $I$ appearing in (\ref{eq:offdiag}) emerges from the process
with an incoming virtual photon, two pions in the loop and two outgoing pions.
In contrast, the loop function appearing in (\ref{eq:BS}) emerges from the
process with two incoming pions instead of the virtual photon. One could
imagine that in principle both processes are renormalized in different 
ways.
Thus we feel legitimated to explore the consequences of 
different renormalization points $\mu$ and $\tilde \mu$. Nonetheless, we recall
the argument from \cite{Lutz:2003fm} that the choice (\ref{eq:same-renorm})
ensures approximate crossing symmetry. We will come back to that point
below.

The pion form-factor emerges from the scattering 
amplitude $t_{e^+ e^- \to \pi^+ \pi^-}$ by normalizing to the QED process,
i.e.\ the coupling of the virtual photon to the charge of the pion, 
cf.\ (\ref{eq:KFF}):
\begin{eqnarray}
F_\pi(s)  & = & 
	\frac{t_{e^+ e^- \to \pi^+ \pi^-}(s)}%
		{k^{\rm QED}_{e^+ e^- \to \gamma^* \to \pi^+ \pi^-}(s)} 
\nonumber \\ 
	&	= &
	\frac{k_{e^+ e^- \to \pi^+ \pi^-}(s)}%
 		{k^{\rm QED}_{e^+ e^- \to \gamma^* \to \pi^+ \pi^-}(s)} \,
		\left[ 1 + \left(I(s) - {\rm Re} \, I(\tilde \mu^2) \right) \, t_{l=1,I=1}(s) 
		\right]  
\nonumber \\
	&	= &
	\left( 1 + 
		\frac{k^{\rm res}_{e^+ e^- \to \gamma^* \to \rho^0 \to \pi^+ \pi^-}(s) }%
			{k^{\rm QED}_{e^+ e^- \to \gamma^* \to \pi^+ \pi^-}(s) } 
	\right) \,
	\left[ 1 + \left(I(s) - {\rm Re} \, I(\tilde \mu^2) \right) \, t_{l=1,I=1}(s) 
		\right] 
\nonumber \\
	&	=: & 
	F^{\rm tree}_\pi(s) \, 
	\left[ 1 + \left(I(s) - {\rm Re} \, I(\tilde \mu^2) \right) \, t_{l=1,I=1}(s) 
		\right]  \,.
\label{eq:pionFF-calc}
\end{eqnarray}
With the conventions of \cite{Lutz:2008km} and ignoring the electron mass one gets
\begin{equation}
F^{\rm tree}_\pi(s) = 1 - \frac{m_V	^2 \, h_P \, e_V}{16 \, e \, f^2} \,
\frac{s}{s-m_{\rho,{\rm bare}}^2}
\label{eq:pionFF-tree}
\end{equation}
with the electromagnetic charge $e$ of the pion $\pi^+$. The dimensionless 
quantity $e_V$ parametrizes the coupling of the photon to the 
elementary vector meson.

We have argued above that it is reasonable to explore the consequences of choosing 
renormalization points $\mu$ and $\tilde \mu$ different from each other and different
from \eqref{eq:same-renorm}. It is important to add, however, that this line of reasoning 
does not hold for processes with an elementary
vector meson. In this case the pion loop starts at the three-point vertex with the
vector meson, no matter whether the vector meson has emerged from a photon or from
incoming pions. Thus one has to renormalize always the same type of loop.
Consequently, for the scenario with an elementary vector meson one should keep
at least $\mu = \tilde \mu$. Also from a technical point of view the equality
of $\mu$ and $\tilde \mu$ is necessary: The tree-level singularity of 
$F^{\rm tree}_\pi(s)$ at $s=m_{\rho,{\rm bare}}^2$ is only dressed for the full
$F_\pi(s)$ if both renormalization points coincide. Only in this case one finds:
\begin{eqnarray}
  \label{eq:pionFF-same-ren}
  F_\pi(s)  & = & F^{\rm tree}_\pi(s) \, 
	\left[ 1 + \left(I(s) - {\rm Re} \, I(\mu^2) \right) \, t_{l=1,I=1}(s) 
		\right]
  = \frac{F^{\rm tree}_\pi(s)}{k_{l=1,I=1}(s)} \, 
  \frac{1}{k^{-1}_{l=1,I=1}(s) - \left(I(s) - {\rm Re} \, I(\mu^2) \right)} \,.
\end{eqnarray}
The tree-level pion form-factor $F^{\rm tree}_\pi(s)$ and the 
tree-level pion-scattering amplitude $k_{l=1,I=1}(s)$ both
diverge at $s=m_{\rho,{\rm bare}}^2$, but their ratio remains finite. 
The rewriting \eqref{eq:pionFF-same-ren} cannot be achieved for $\mu \neq \tilde \mu$.

To summarize the renormalization issue: For the scenario with an elementary vector
meson we have only one renormalization point. To ensure approximate crossing symmetry
we will use \eqref{eq:same-renorm}, but explore moderate deviations from this relation.
For the scenario without an elementary vector meson we will study the consequences
of arbitrary choices for the two renormalization points.

\subsection{Resummations in the Bethe-Salpeter and the Dyson-Schwinger equation}
\label{sec:spec}

\begin{figure}
  \centering
  \includegraphics[keepaspectratio,width=0.45\textwidth]{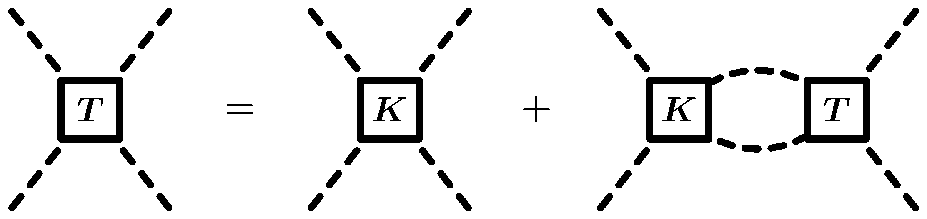}  \\[2em]
  \includegraphics[keepaspectratio,width=0.4\textwidth]{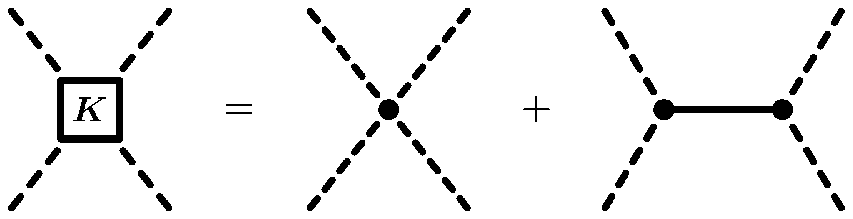}
  \caption{Bethe-Salpeter equation (top) and its kernel (bottom) 
    from \eqref{eq:res-nonres-gen-k}. The dashed lines correspond to the scattering
    partners, the full line to the (bare) resonance propagator.}
  \label{fig:fig-BS}
\end{figure}

\begin{figure}
  \centering
  \includegraphics[keepaspectratio,width=0.45\textwidth]{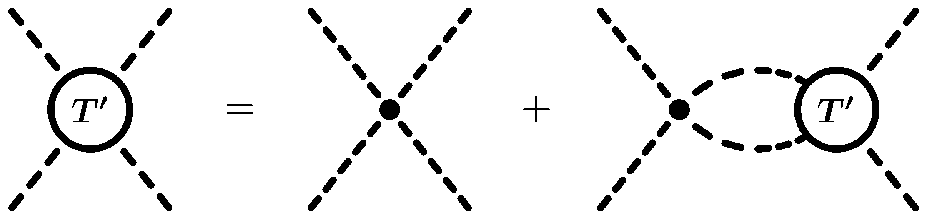}
  \caption{Bethe-Salpeter equation for a purely point-like kernel.}
  \label{fig:fig-nonres}
\end{figure}

The discussion in the present subsection will concern, of course, only the
scenario with an elementary vector meson. We will start out slightly more
general than the case considered in the rest of the paper. At the same time
we will be somewhat more schematic (e.g., by concentrating on scalar quantities for the 
Dyson-Schwinger equation and disregarding renormalization issues).
Suppose that a scattering process happens via a resonant and a non-resonant
subprocess, i.e.\ the scattering kernel is given by 
\begin{equation}
  \label{eq:res-nonres-gen-k}
  K = g_1^2(s) - \frac{g_2^2(s)}{s-m_{\rm bare}^2} = 
  g_1^2(s) - g_2^2(s) \, D_{\rm bare}(s)  \,,
\end{equation}
cf.\ \eqref{eq:k-scatt-tree}. Here $g_1^2$ can be regarded as coming from a point
interaction between the scattering partners while $g_2$ is the coupling of the
scattering partners to the resonance. The scattering amplitude as obtained from
the Bethe-Salpeter equation, cf.\ \eqref{eq:BS}, \eqref{eq:def-loop}, is given by
\begin{equation}
  \label{eq:gen-t}
  T = \frac{K}{1-K \, I}  \,.
\end{equation}
The graphical version is displayed in Fig.\ \ref{fig:fig-BS}.
For convenience we also introduce the modified scattering amplitude which emerges
if there was only the non-resonant part in the kernel:
\begin{equation}
  \label{eq:gen-tprime}
  T' = \frac{g_1^2}{1- g_1^2 \, I}  \,.
\end{equation}
For illustration we refer to Fig.\ \ref{fig:fig-nonres}.

\begin{figure}
  \centering
  \includegraphics[keepaspectratio,width=0.3\textwidth]{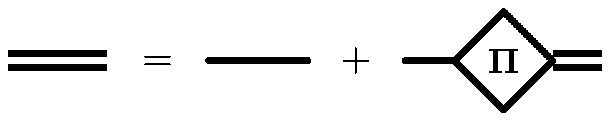}
  \caption{The Dyson-Schwinger equation. The full line denotes the bare and the double 
    line the full resonance propagator.}
  \label{fig:fig-DS}
\end{figure}

Next we turn to the Dyson-Schwinger equation. It is given by
\begin{equation}
  \label{eq:Dyson}
  D_{\rm full}^{-1} = D_{\rm bare}^{-1} - \Pi
\end{equation}
and displayed in Fig.\ \ref{fig:fig-DS}. 
All non-trivial information is contained in the self energy $\Pi$ which appears in
\eqref{eq:Dyson}. 
The standard approach is to approximate the self energy by the one-loop expression. For our
case at hand this is given by
\begin{equation}
  \label{eq:pi-1loop}
  \Pi_{\rm one \, loop} = -g_2^2 \, I  \,.
\end{equation}
Within the same spirit one calculates the scattering amplitude in resonance
approximation:
\begin{equation}
  \label{eq:t-resapprox}
  T \stackrel{?}{\approx} T_{\rm res} := - g_2^2 \, D_{\rm full}  \,.
\end{equation}

Of course, in the absence of the point interaction $\sim g_1^2$ in 
\eqref{eq:res-nonres-gen-k} --- or in practice if all non-resonant interactions
are sufficiently small --- the result of the Bethe-Salpeter equation is identical
to the result of \eqref{eq:t-resapprox}, if the full propagator is obtained from the
Dyson-Schwinger equation \eqref{eq:Dyson} within the one-loop approximation 
\eqref{eq:pi-1loop} for the self energy. In this case one gets
\begin{eqnarray}
  \label{eq:t-g10}
  \left. T \; \vphantom{\int}\right\vert_{g_1^2 =0} & = & 
  \left. \frac{1}{K^{-1} - I} \, \right\vert_{g_1^2 =0} 
  = \frac{-g_2^2}{D_{\rm bare}^{-1} + g_2^2 \, I} = - g_2^2 \, D_{\rm full} 
  = T_{\rm res}  \,.
\end{eqnarray}

\begin{figure}
  \centering
  \includegraphics[keepaspectratio,width=0.8\textwidth]{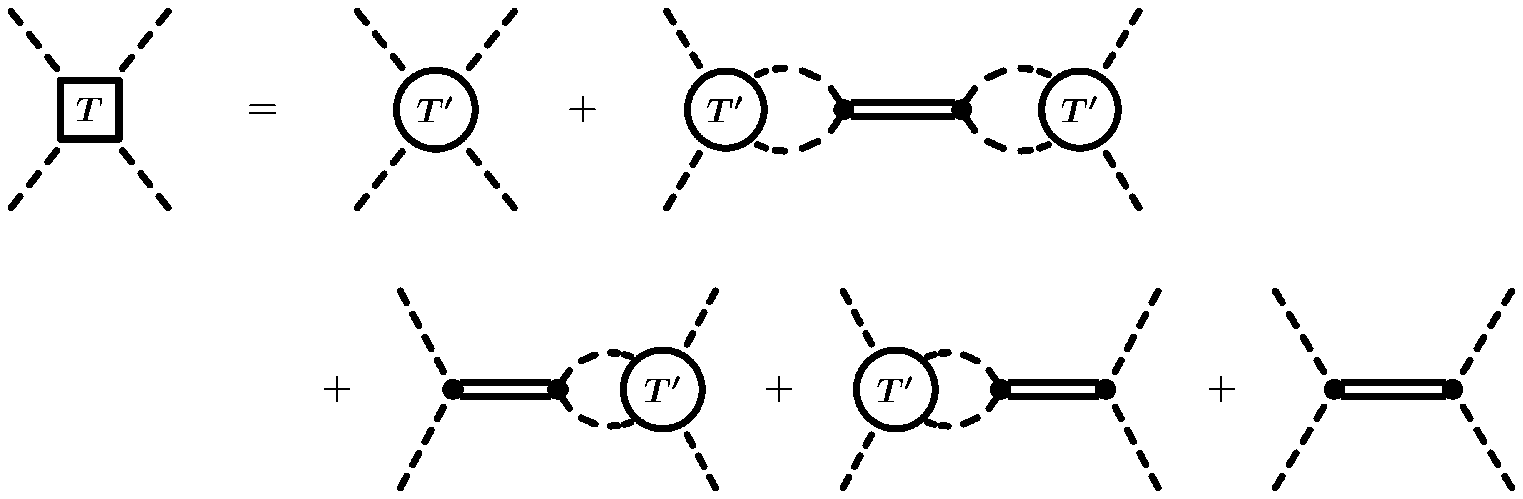}
  \caption{Representation of the scattering amplitude via the full propagator and the
    non-resonant terms. For details see the captions of Figs.\ \ref{fig:fig-BS}
    and \ref{fig:fig-DS}.}
  \label{fig:BS-DS}
\end{figure}

\begin{figure}
  \centering
  \includegraphics[keepaspectratio,width=0.4\textwidth]{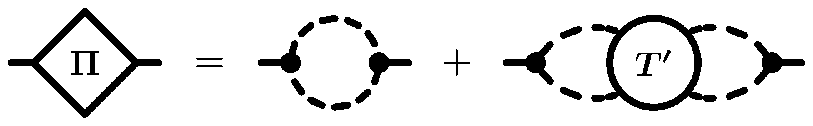}
  \caption{The improved self energy. 
    For details see the captions of Figs.\ \ref{fig:fig-BS}
    and \ref{fig:fig-DS}.}
  \label{fig:self-impr}
\end{figure}

However, for the more general case $g_1^2 \neq 0$ it should be clear that the result 
from the Bethe-Salpeter equation \eqref{eq:gen-t} 
resums more processes than the one-loop plus resonance approximation, \eqref{eq:pi-1loop} 
and \eqref{eq:t-resapprox}, respectively. Therefore we ask how we have to improve
\eqref{eq:pi-1loop} and \eqref{eq:t-resapprox} such that they contain the same information
as \eqref{eq:gen-t}. The demand is that the scattering amplitude can be obtained from
the full propagator plus non-resonant terms. It is natural to expect the following relation
\begin{equation}
  \label{eq:t-full-prop}
  T = T' - T' \, I \, g_2 \, D_{\rm full} \, g_2 \, I \, T'
  - g_2 \, D_{\rm full} \, g_2 \, I \, T'
  - T' \, I \, g_2 \, D_{\rm full} \, g_2 
  - g_2 \, D_{\rm full} \, g_2  \,,
\end{equation}
which is graphically displayed in Fig.\ \ref{fig:BS-DS}. 
Note that here the modified
scattering amplitude \eqref{eq:gen-tprime} shows up which resums the non-resonant 
scattering kernel. It is a straightforward exercise to show that \eqref{eq:t-full-prop}
is equivalent to \eqref{eq:gen-t} if the self energy is given by
\begin{equation}
  \label{eq:pi-improved}
  \Pi = -g_2^2 \, I  - g_2^2 \, I^2 \, T' = -g_2^2 \, I \, \frac{1}{1-g_1^2 \, I}  \,.
\end{equation}
The graphical version of this relation is shown in Fig.\ \ref{fig:self-impr}. 
Note
that the appearance of $T'$ instead of $T$ in \eqref{eq:pi-improved} is very natural:
The self energy contains only one-particle irreducible diagrams. This would be spoiled
by the appearance of any intermediate resonance propagator. $T$ contains such diagrams
while $T'$ does not. 

The physics point we want to make is the following: In resonance models one
typically uses \eqref{eq:t-resapprox} and \eqref{eq:pi-1loop}. However, from a
more general effective-field-theory point of view one should not disregard point
interactions --- or more general non-resonant interactions --- without checking
their importance. Here the Bethe-Salpeter equation provides a natural tool.
Alternatively, one can use the Dyson-Schwinger equation together with the
improved relations \eqref{eq:t-full-prop} and \eqref{eq:pi-improved}. Even if
one includes point interactions it is common practice to treat them on a perturbative
level while for the resonant part a Dyson-Schwinger resummation is performed. The
reason is clear from a practical point of view: A bare resonance propagator
needs a width to provide sensible, non divergent results. For the non-resonant
interactions it is not obvious that something is missing if one does not resum them.
However, from the point of view of exact unitarity one should resum all interactions, 
just like the Bethe-Salpeter equation automatically does.

Finally it is interesting to study how the high-energy behavior is modified when
changing from the one-loop self energy \eqref{eq:pi-1loop} to the improved version
\eqref{eq:pi-improved}. For point interactions $g_1^2(s)$ is a polynomial in $s$ and is
real. The same holds for $g_2(s)$. 
After renormalization the loop function $I(s)$ diverges logarithmically with $s$
(cf.\ \eqref{eq:defJ}). Its imaginary part reaches a constant. Therefore the
real part of the quantity $\Pi_{\rm one \, loop}/g_2^2$ diverges logarithmically while
its imaginary part becomes constant. In contrast, the improved quantity
$\Pi/g_2^2$ has a real part which converges $\sim 1/g_1^2$ and an imaginary part
which drops to zero $\sim 1/(g_1^4 \, \log^2 s)$. 

After these general considerations
how a Bethe-Salpeter resummation compares to a Dyson-Schwinger resummation
we return to the main subject of the present work. We recall that the task is
to figure out whether one needs an elementary vector meson to describe the experimental
data for the electromagnetic pion form-factor or whether it is sufficient
to iterate the point interaction of lowest-order chiral perturbation theory.

\section{Results}
\label{sec:results}

The pion form-factor is given in \eqref{eq:pionFF-calc}. With the ingredients
specified in \eqref{eq:pionFF-tree}, \eqref{eq:renorm}, \eqref{eq:defJ},
\eqref{eq:BS} and \eqref{eq:k-scatt-tree} one obtains an expression which in
general depends on the resonance parameters $h_P$, $e_V$ and $m_{\rho,{\rm bare}}$ 
of the bare vector meson and on the renormalization points $\mu$ and $\tilde \mu$. In the
following we will discuss the two scenarios where an elementary (bare) vector meson
is not included or is included, respectively.

\subsection{Scenario without an elementary vector meson}
\label{sec:wo}

\begin{figure}
  \centering
  \includegraphics[keepaspectratio,width=0.7\textwidth]{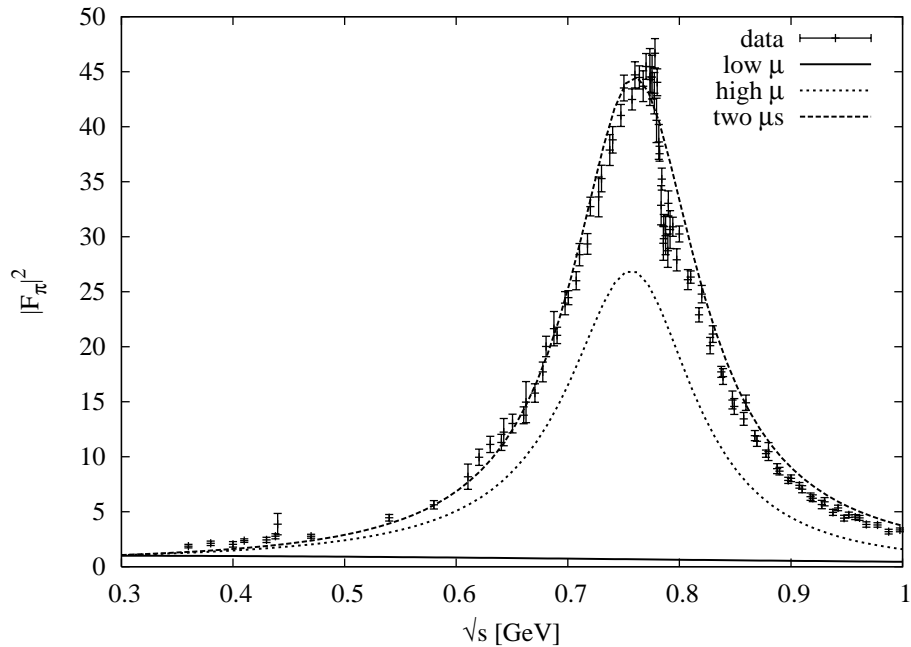}
  \caption{Description of the modulus square of the pion form-factor 
    in the Bethe-Salpeter approach
    without an elementary vector meson, i.e.\ including in the kernels only the 
    interactions from lowest-order chiral perturbation theory.
    The full line, labeled with ``low $\mu$'', 
    denotes the physically reasonable case where the renormalization points
    are chosen according to \eqref{eq:same-renorm}. The other lines denote the technically
    possible, but physically questionable cases where the renormalization points are chosen
    in the TeV range. The dotted line, labeled with ``high $\mu$'', is obtained for
    $\mu = \tilde\mu = 1.1 \,$TeV and the dashed line, labeled with ``two $\mu$s'', for
    $\mu = 1.1 \,$TeV, $\tilde \mu = 10 \,$TeV. See main text for details.
    Data taken from \cite{Barkov:1985ac,Akhmetshin:2001ig,Akhmetshin:2003zn}.}
  \label{fig:norho}
\end{figure}

The case where there is no elementary vector meson is easily obtained by setting
$h_P =0$ in \eqref{eq:k-scatt-tree}, \eqref{eq:pionFF-tree}. 
In this case the values for $e_V$ and $m_{\rho,{\rm bare}}$ are irrelevant. The only
parameters left are the renormalization points. Demanding in addition approximate
crossing symmetry for the scattering amplitudes resulting from the Bethe-Salpeter 
equation, i.e.\ fixing the renormalization points according to \eqref{eq:same-renorm},
one is left with a parameter-free result. This result is compared to data in 
Fig.\ \ref{fig:norho}, full line labeled with ``low $\mu$''. 
Obviously, the pion four-point interaction of leading-order chiral perturbation
theory, iterated by the Bethe-Salpeter equation, is insufficient to create the peak
seen in the data. This confirms the findings of \cite{Oller:1998zr}. Hence, one needs
in addition an elementary vector meson and in the following subsection we will demonstrate
that then an excellent description of the pion form-factor can be obtained.

The physics case is closed with the previous remarks. 
Nonetheless, it is interesting to explore whether it is {\em technically} 
possible at all to
come close to the data in an approach without an elementary vector meson. Therefore, we
abandon the constraints \eqref{eq:same-renorm} in the following and study the dependence
of our result on the renormalization points $\mu$ and $\tilde \mu$. It turns out that
for choices below the TeV range no appreciable peak is obtained. 
(We also refer to \cite{Oller:1998zr} for a similar discussion concerning the 
pion-scattering amplitude.)
If one keeps the two renormalization points the same, one can generate a peak for large 
values of $\mu = \tilde \mu$, but
not a decent description of the data. An example is provided in Fig.\ \ref{fig:norho}
by the dotted line labeled with ``high $\mu$''. 
A (technically but not physically)
satisfying description of the data is obtained, if the renormalization points are
allowed to differ from each other. This is demonstrated in Fig.\ \ref{fig:norho}
by the dashed line labeled with ``two $\mu$s''. Here we have chosen
$\mu = 1.1 \,$TeV, $\tilde \mu = 10 \,$TeV. 

The technical success of this 
approach is nicely explained in \cite{Hyodo:2008xr}: It is shown there that an improper
choice of the renormalization point(s), i.e.\ a choice which deviates significantly
from \eqref{eq:same-renorm}, mimics an elementary resonance. Therefore, such an improper
choice can lead to the wrong conclusion. The naive interpretation of the good agreement
between the dashed line in Fig.\ \ref{fig:norho} and the data would be, that one can
describe the pion form-factor without an elementary vector meson. The correct
interpretation, however, is that one has introduced an elementary resonance through
the backdoor by the (improper) choice for the renormalization points.

\subsection{Scenario including an elementary vector meson}
\label{sec:with}

The previous considerations also tell us something for the case to which we turn now, 
namely
the one {\em with} an elementary vector meson. Since an improper choice for the
renormalization points mimics an elementary resonance, one generates double counting
in such a case. Both the included elementary state and the wrongly chosen renormalization
point would introduce one and the same resonance. This is avoided by the constraint
\eqref{eq:same-renorm} which we use from now on more or less. 
Still we will study the consequences of small deviations from \eqref{eq:same-renorm}.

\begin{figure}
  \centering
  \includegraphics[keepaspectratio,width=0.7\textwidth]{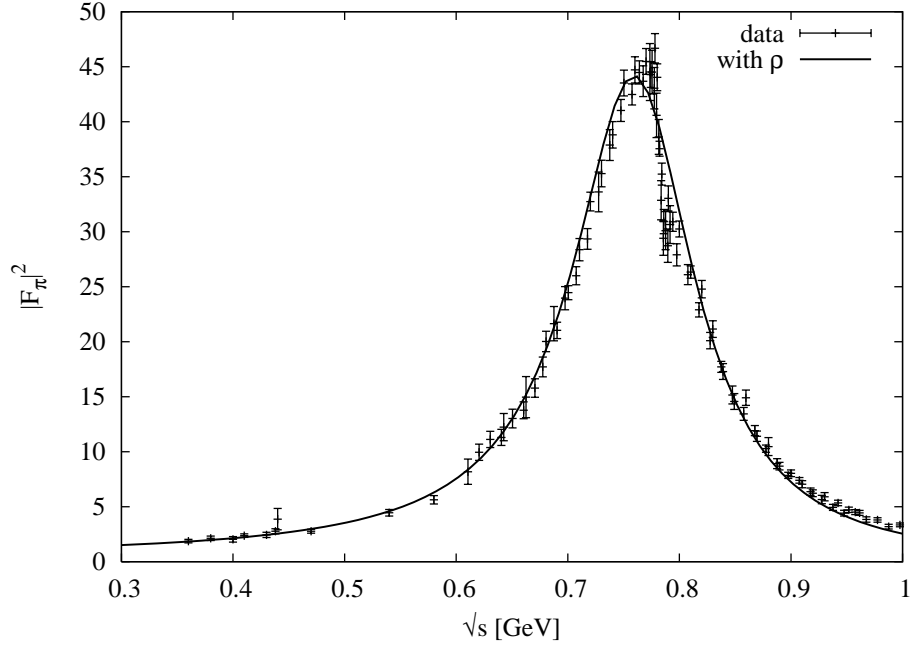}
  \caption{Description of the modulus square of the pion form-factor 
    in the Bethe-Salpeter approach
    including in the kernels 
    an elementary vector-meson resonance
    together with the interactions from lowest-order chiral perturbation theory.
    The resonance parameters are chosen such that
    a good description of the pion form-factor is obtained
    concerning height (mainly
    controlled by parameter $e_V$), peak position ($m_{\rho,{\rm bare}}$) 
    and width ($h_P$). The renormalization points
    are chosen according to \eqref{eq:same-renorm}. See main text for details.
    Data taken from \cite{Barkov:1985ac,Akhmetshin:2001ig,Akhmetshin:2003zn}.}
  \label{fig:withrho}
\end{figure}

Having fixed the renormalization points, we are left with three parameters:
The coupling of the virtual photon to the elementary vector meson $\sim e_V$,
the coupling of the pions to this  vector meson $\sim h_P$, and the (bare) mass
of this vector meson, $m_{\rho,{\rm bare}}$. It should not be too surprising that
these parameters can be used to fix the height, the position and the width of the peak
and obtain in this way a good description of the data. This is demonstrated in
Fig.\ \ref{fig:withrho}. We note in passing that the largest deviation of our curve
from the data happens at the small additional narrow peak slightly to the right of the
peak position of the broad main peak. This is just the appearance of the $\omega$ meson
due to the already mentioned isospin violating $\rho$-$\omega$ mixing. 
We have chosen the following parameters to produce the theory curve displayed in
Fig.\ \ref{fig:withrho}: 
\begin{eqnarray}
  \label{eq:respara}
  h_P = 0.304 \,, \qquad e_V = 0.228 \,, \qquad m_{\rho,{\rm bare}} = 0.711 \,\mbox{GeV.}
\end{eqnarray}
The parameters $h_P$ and $e_V$ have also been determined in 
\cite{Lutz:2008km} from the partial decay widths of the two-body decays of vector mesons. 
Our results \eqref{eq:respara} are more precise, but in full agreement with the
determination of \cite{Lutz:2008km}. The precise value for $h_P$ from \eqref{eq:respara} 
has already been used in \cite{Leupold:2008bp} with good success in the description
of the three-pion decay of the $\omega$ meson.

\begin{figure}
  \centering
  \includegraphics[keepaspectratio,width=0.7\textwidth]{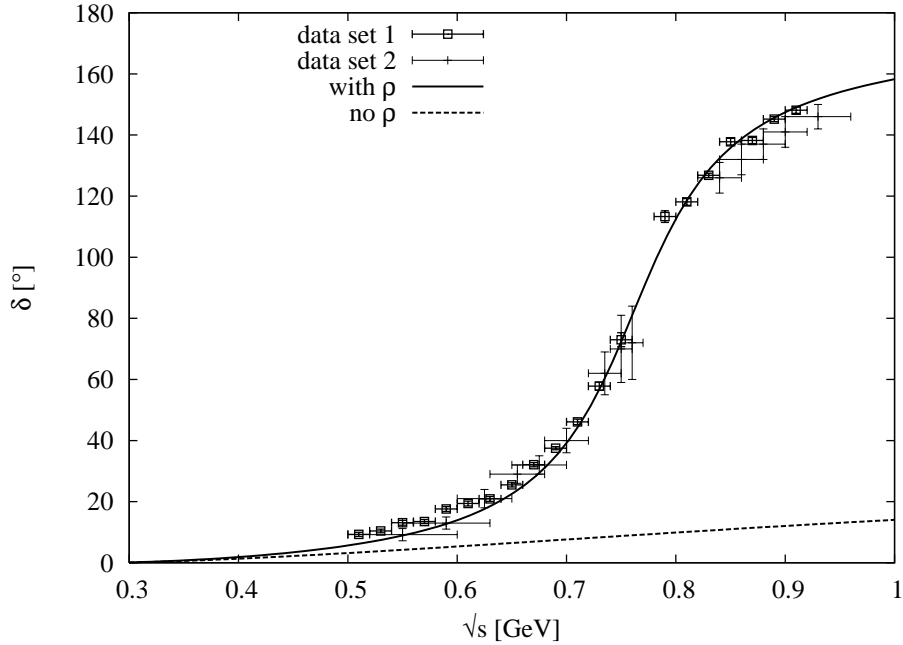}
  \caption{Description of the pion-scattering phase shift in the Bethe-Salpeter approach.
    Dashed line labeled with ``no $\rho$'': including in the kernel only
    the interaction from lowest-order chiral perturbation theory.
    Full line labeled by ``with $\rho$'': including in addition
    an elementary vector-meson resonance.
    The resonance parameters are chosen such that
    a good description of the pion form-factor is obtained, cf.\ Fig.\ \ref{fig:withrho}.
    See main text for details.
    Data set 1/2 taken from \cite{Estabrooks:1974vu}/\cite{Protopopescu:1973sh}.}
  \label{fig:ps}
\end{figure}

For completeness we also show the pion-scattering phase shift introduced 
in \eqref{eq:l-el-ph-sh}. The result is displayed in Fig.\ \ref{fig:ps}.
Note that our intention was to achieve a good description of the pion form-factor. Hence
our resonance parameters \eqref{eq:respara} have not been tuned to describe the
pion-scattering phase shift. In view of that we can be rather satisfied with the
full line of Fig.\ \ref{fig:ps} which uses \eqref{eq:respara} and \eqref{eq:same-renorm}.
We have also displayed the result for the phase shift if no elementary vector meson
is included using  \eqref{eq:same-renorm}; dashed line in Fig.\ \ref{fig:ps}. 

Finally we explore the consequence of a small deviation from \eqref{eq:same-renorm}:
Varying $\mu = \tilde \mu$ between the electron mass (i.e.\ essentially zero)
and twice the pion mass and keeping the resonance parameters 
unchanged from \eqref{eq:respara} does not lead to 
visible changes in Fig.\ \ref{fig:withrho}. Even for larger deviations from 
\eqref{eq:same-renorm} one can argue that a change in the renormalization point can
be largely compensated by a change in the resonance parameters. After all, what 
determines the pion form-factor to a large extent is the elementary vector meson.
In other words, the pion form-factor is more or less fixed by its total height,
the peak position and the peak width. These quantities are dominantly influenced
by the vector-meson--photon coupling $e_V$, 
the bare vector-meson mass $m_{\rho,{\rm bare}}$
and the vector-meson--pion coupling $h_P$, respectively.

\acknowledgments The author thanks H.\ van Hees for discussions and 
for reading the manuscript and 
U.\ Mosel for continuous support. This work is supported by GSI Darmstadt. 

\appendix

\section{Two-channel Bethe-Salpeter equation 
  with a perturbative channel}

In this appendix we keep the discussion more general than in the rest of the paper, 
but also more schematic. We consider a two-channel Bethe-Salpeter equation
\begin{equation}
T^{-1} = K^{-1} - I
\label{eq:BS-2chan}
\end{equation}
with symmetric $2 \times 2$ matrices $T$, $K$ and $I$ where $I$ is a diagonal matrix.
We want to determine the off-diagonal matrix element $T_{12}$ for the case that 
all matrix elements of $K$ except for $K_{11}$ are very small. A straightforward
exercise yields
\begin{eqnarray}
T_{12} & = & 
\frac{K_{12}}{1-I_{11}K_{11}-I_{22}K_{22}-I_{11}I_{22}(K_{12}^2-K_{11}K_{22})}
\nonumber \\
& \approx &
\frac{K_{12}}{1-I_{11}K_{11}}
\approx
K_{12} \, \left( 1 + I_{11}T_{11} \right)  \,.
\label{eq:t12approx}
\end{eqnarray}
In the last step we have used the fact that $T_{11}$ can be obtained in 
lowest-order approximation from the one-channel Bethe-Salpeter equation
\begin{equation}
T_{11}^{-1} \approx K_{11}^{-1} - I_{11}  \,,
\label{eq:BS-onech}
\end{equation}
if all other entries of $K$ are perturbatively small:
\begin{eqnarray}
T_{11} & = & 
\frac{K_{11}+I_{22}(K_{12}^2-K_{11}K_{22})}{1-I_{11}K_{11}-I_{22}K_{22}-I_{11}I_{22}(K_{12}^2-K_{11}K_{22})}
\nonumber \\
& \approx &
\frac{K_{11}}{1-I_{11}K_{11}}  \,.
\label{eq:t11approx}
\end{eqnarray}

\bibliography{literature,literature2}
\bibliographystyle{apsrev}

\end{document}